# Density functional theory study of the structural, electronic, and surface reaction properties of bismuth vanady1 oxyhalide BiVO$_3$F

Taifeng Liu*, Tongling Liu, and Yongqiang Zheng

National & Local Joint Engineering Research Center for Applied Technology of Hybrid Nanomaterials, Henan University, Kaifeng 475004, China

Corresponding author: Taifeng Liu: tfliu@vip.henu.edu.cn

Abstract:

BiVO$_3$F is a promising material used in solar energy conversion systems. Here, we first report the calculated structural, electronic, and surface reaction properties using PBE and hybrid density functionals. We found it is a direct band gap semiconductor, and the calculated band gap is consistent with experimental value only using the hybrid density functional with a fraction of Hartree Fock (HF) exchange α=0.1. The (001) surface is the most stable surface among all the low index (001), (010), and (100) surfaces. There are V and Bi sites exposed on (001) surface which can serves as activity sites. That is quite different from BiVO$_4$ where only Bi sties can be taken as surface reaction sites. The OER intermediates OH* and OOH* prefer to form a bridge structure on both V and Bi sites. This makes the first proton removal step is very easy, but the O-O bond is difficult to form which leads the overpotential of OER is very high. Our work plays a guide principle to design the high efficiency photocatalysis and photoanodes based on BiVO$_3$F.

1. Introduction

Semiconductors based photocatalytic (PC) and photoelectrocatalytic (PEC) water splitting has been considered a promising route to convert the sunlight into fuels and solve energy and environmental crises[1]. There are mainly three steps in PC or PEC water splitting which are light absorption, charge separation and transport, and surface reactions. Among all the semiconductors, BiVO$_4$ (BVO) has attracted much attention in recent years because of its visible-light absorption, suitable conduction and valence band edge positions, and low-cost facile synthesis route[2]. However, BVO is still suffering from low carrier separation and transport efficiency and slow surface oxygen evolution kinetics[3]. To overcome these weaknesses, various modification strategies, including morphology control, element doping, heterostructures, plasmonic enhancement and surface passivation, have been proposed and implemented to improve its PC or PEC activity. Fluorination of the metal oxide is also a significant modification strategy to increase the photocatalytic efficiency[4].

Recently, the first bismuth vanady1 oxyhalide BiVO$_3$F (BVOF) was synthesized by Mentre et al.[5]. Comparing to the most promising photoanode BiVO$_4$ (BVO), the introduction of d$^1$ V$^{4+}$ species has several advantages. It is well known that in the BVO, the excess electrons will be localized on the V site to form small polarons which leads the slow charge transport[6-11]. The d$^1$ V$^{4+}$ species in BVOF will affect the excess electrons localization on V site to form small polarons and improve the charge separation and transport. The direct d-d excitations of the d$^1$ V$^{4+}$ species response for a narrowed bandgap with a value ~ 1.7 eV in the BVOF while in BVO the bandgap is ~ 2.4 eV. The light absorption will be enhanced which is benefit for the photocatalysis. Moreover, in the BVOF, the V atoms will be exposed as activity sites, while in BVO the VO4 tetrahedron is rigid and only Bi atoms serves as activity sites[12].

As far as we know, there are no theoretical investigations on BVOF. In this work, we first report the electronic structure and surface reaction calculations using first principal methods. We found the bandgap of BVOF is consistent with the experimental result only using the PBE functional with a fraction of Hartree Fock (HF) exchange α=0.1. The (001) surface has the lowest surface energy and both V and Bi atoms can service as activity sites for oxygen evolution reaction (OER). The overpotential of OER is high on this surface as the O-O bond is difficult to form during the OER process.

The paper is organized as following: in section 2, we give the computational details, in section 3, we present the results and discussion, and the conclusion is given in section 4.

2. Computation Details

We performed spin-polarized DFT calculations using the Vienna ab initio Simulation Package (VASP).[13, 14] For structure determination we adopted the Perdew-Burke-Ernzerh (PBE) parametrization of the exchange and correlation potential in the generalized gradient approximation (GGA)[15].The projector-augmented wave method was applied to describing electron–ion interactions[16, 17]. The structures were relaxed until all forces on atoms were less than 0.01 eV Å$^{-1}$, and a plane-wave cutoff energy of 600 eV was employed. Self-consistent DFT energies were converged to 10$^{-4}$ eV. The core electrons were represented with pseudopotentials, with 6, 7, 11 and 15 valence electrons left in plane wave description for O, F, V and Bi, respectively. The geometry of the primitive cell of BVOF was optimized using PBE functional, and the Γ-centered k-point mesh of the Brillouin zone sampling was set at 7 × 7× 3 based on the Monkhorst−Pack scheme.[18] After optimization, the electronic structure was calculated using hybrid functional, and the k-points was set at 2 × 2× 1. The 4 × 4× 2 k-points was also used to check the electronic structure.

We calculated the phonon frequencies of BVOF at gamma point with VASP code itself. To do so, in the geometry optimization, the cutoff energy of 900 eV, all forces on atoms less than $10^{-4}$ eV Å$^{-1}$, and self-consistent DFT energies converged to $10^{-8}$ eV were employed.

The low index (100), (010), and (100) surfaces was considered in this work. The symmetric and stoichiometric slabs were constructed for these surfaces with a vacuum layer of 20 Å was used in order to avoid the interaction between periodic slabs. There are 14, 8 and 6 Bi layers in (100), (010), and (100) slabs, respectively. The 16, 10 and 8 Bi layers of the three slabs was also considered to check the thickness of the slabs. For the surface energies calculations, the primitive cell of the slab was used, and the k points were set at 7 × 3× 1, 3 × 7× 1, and 7 × 7× 1 for (100), (010), and (001) slabs, respectively. For the OER calculations, only a 2 × 2× 1 supercell of (001) slab was used, and the k-points was reduced to 2 × 2× 1. The surface formation energies $\gamma$ are calculated using following equation:

$$\gamma = \frac{E_{slab} - mE_{bulk}}{2A} \qquad (1)$$

where $E_{slab}$ and $E_{bulk}$ are the total energies of the slabs and bulk, $m$ is the number of the BVOF units in the slabs, and $A$ is the total exposed area of the two identical sides of the slab.

For the OER calculations, we followed Nørskov's four step PCET protocol[19] with intermediates denoted *, OH*, O*, and OOH* throughout, and the OER over-potential determined by:

$$E_{OP}(V) = \frac{\max_{n=1,4}\{\Delta G_n\}}{e} - 1.23 \; [V] \qquad (2)$$

where $\Delta G_n$ in eV denotes the step free energy for *step n* ($n$ = 1 to 4), with the zero of energy taken as the energy of the free surface site (denoted *) plus the energy of two water molecules, $n$ is the step index (1 to 4), $e$ is the electron charge. The rate-limiting step is the step giving rise to the value $E_{OP}$. Step free energies were calculated from the DFT step energies by applying a contribution for changes in zero-point energy (ZPE) and entropy (TS) for the various steps (given below). ZPE and TS corrections were taken from Valdes *et al.*[19] and from Rossmeisl *et al.*[20]

3. Results and Discussion

3.1 The structural properties of BVOF

We take the initial geometry of BVOF from Mentre et al's.[5] experiment. It is a monoclinic structure with the lattice parameters are a=5.262 Å, b=4.972 Å, c=12.615 Å, α=γ=90º, β=95.59º. After optimization with PBE functional, the primitive structure of BVOF is shown in the Figure 1(a), and the lattice parameters become a=5.387 Å, b=5.051 Å, c=12.751 Å, α=γ=90º, β=95.84º which is consistent with experimental values. In Figure 1(b), we show the $2 \times 2 \times 2$ supercell of BVOF which clearly show the S=1/2 chain. The phonon frequencies of BVOF at gamma point are shown in the Table S1 in the supporting information (SI). There are no soft modes at the gamma point, and it indicates it is a stable structure. We also calculated the energy difference between ferromagnetic (FM) and antiferromagnetic (AFM) of BVOF and found that the energy of AFM is only 36 meV lower than the FM structure.

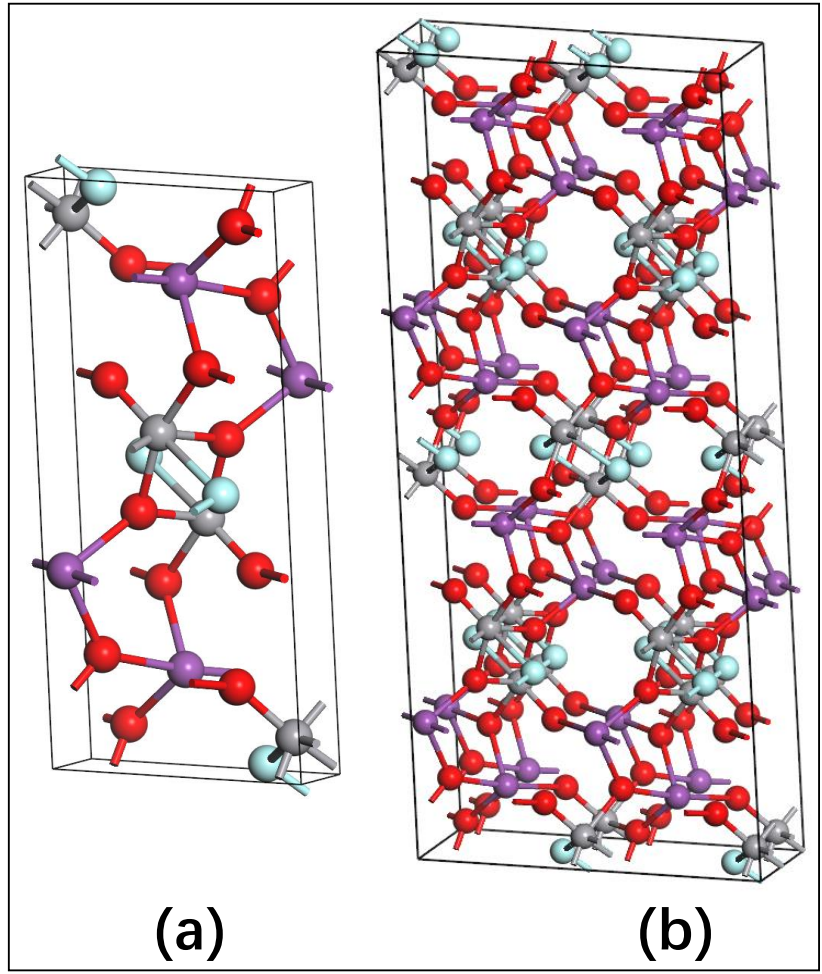

Figure 1. Structure of BVOF, (a) primitive cell, and (b) $2 \times 2 \times 2$ supercell. The purple, grey, red, and light blue spheres are Bi, V, O, and F atoms.

3.2 Electronic structure description using PBE and hybrid functionals.

We first calculated the bandgap of FM BVOF using the PBE functional, and found the bandgap is smaller than the experimental value. To gain a better understanding of the xc functional dependence of the bandgap, we investigate how the inclusion of HF exchange affects the bandgap of FM BVOF. The calculations were performed using hybrid functional with a fraction of HF exchange α=0.1, 0.15, and 0.25. The total density of states (DOS) with different α are shown in Figure 2(a). It shows the bandgap is much larger than the experiment using the PBE0 (α=0.25) functional. This behave is the similar as BVO which the bandgap is largely overestimated with the PBE0 functional. The bandgap of BVOF is only consistent with experiment using the hybrid functional with α around 0.1. The calculated bandgap is ~1.7 eV while in the experiment the value is about 1.5 ~1.7 eV. The total DOS of AFM

BVOF with k-points 2 × 2× 1 and 4 × 4× 2 are shown in Figure S1 in SI. The two DOS are similar indicating the k-points 2 × 2× 1 for electronic structure is accurate enough. The calculated bandgap of AFM BVOF is close to experimental result too.

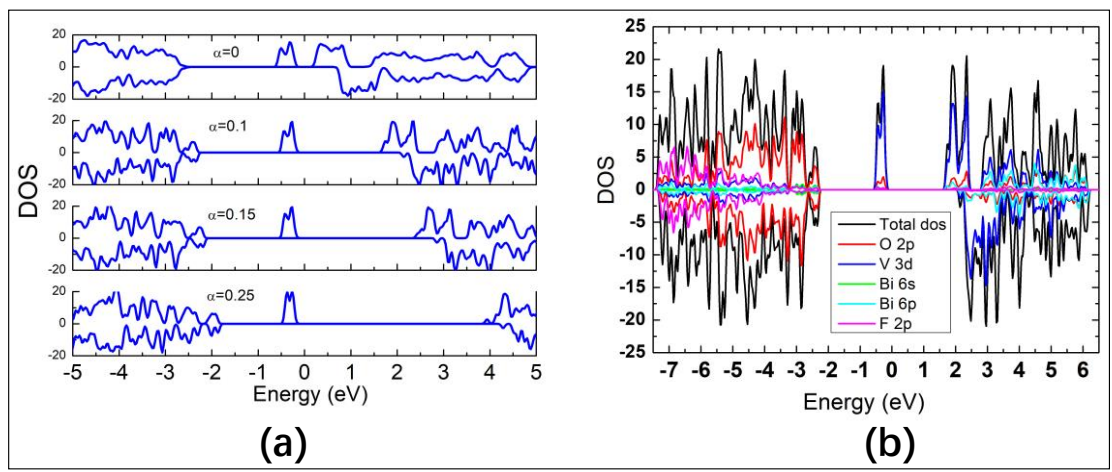

Figure2. (a) The total DOS of BVOF calculated using PBE functional and hybrid functional with different fractions of HF exchange α, (b) the total and projected DOS of BVOF. The Fermi level is set at 0 eV.

The total and projected DOS of FM BVOF calculated using hybrid functional with α=0.1 is shown in Figure 2(b). The valence band (VB) is mainly composed of V 3d states with little contribution of O 2p, F 2p and Bi 6p states. The conduction band (CB) is principally V 3d in character with minor contribution of O 2p and Bi 6p states. The contribution of O 2p, F 2p, and Bi 6s are mainly in the band below VB. When photoexcited, there are mainly V 3d to 3d excitations. There will be holes left on the V sites which will be benefit for the water oxidation as water molecules are mainly adsorbed on the V atoms and the holes can directly transfer to the water and oxide it.

We also calculated the band structure of FM BVOF using PBE functional, and it is shown in Figure S2 and S3 which indicates the indirect gap of FM BVOF.

3.3 Low index (100), (010), and (001) surfaces

For the optimization of slabs and OER intermediates structures, only PBE functional was used as it can describe the geometries correctly. The optimized (100), (010), and (001) slabs are shown in Figure 3. The surface energies of these slabs with different layers are shown in Table 1. The surface energies converged with more layers indicating the thickness of these three slabs are enough. The (001) slab has the lowest surface energy which means it will be easily exposed in the experiment. On the termination of (001) slab, the V and Bi atoms are four and three coordinates while in the bulk there are six and four coordinates, respectively. In the following OER calculations, we only use (001) surface.

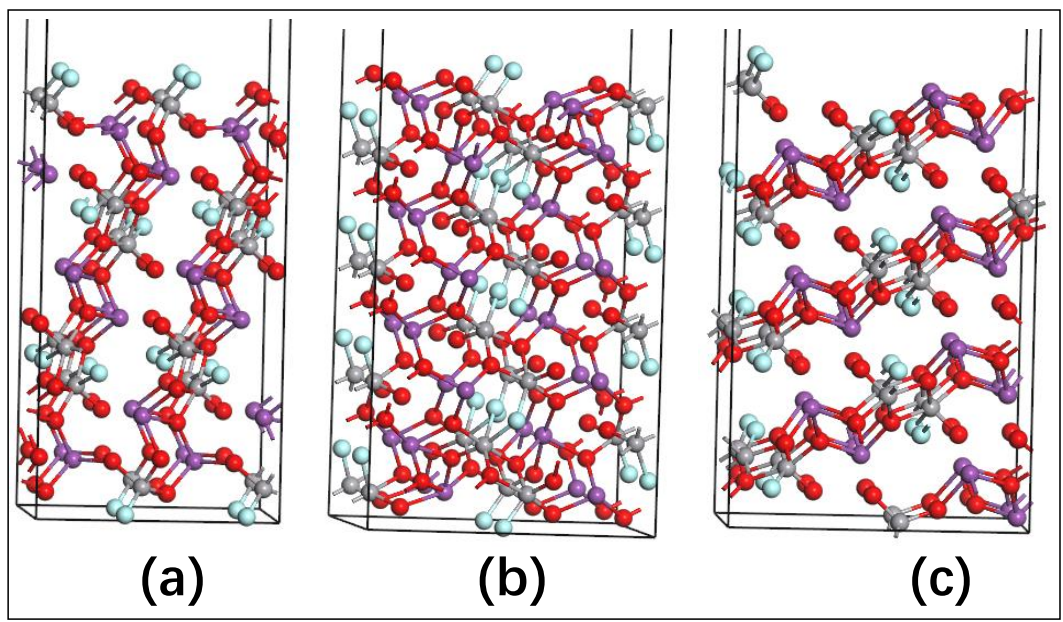

Figure 3. The structures of (a) (001), (b) (010), and (c) (100) slabs.

Table 1. The surface formation energies of (001), (010), and (100) with different layers.

| Slabs | Surface energies (J/m$^2$) | |
|---|---|---|
| | n Layers | (n+2) Layers |
| (100) | 0.7300 | 0.7296 |
| (010) | 0.5603 | 0.5573 |
| (001) | 0.1149 | 0.1137 |

3.4 OER on the (001) surface

On the (100) surface, there are V sites and Bi sites which can be served as OER activity sites. Both sites are considered in our OER calculations. This is different from BVO as only Bi sties exposed on the surface[12]. But after optimization, we found all the OER intermediates OH* and OOH* prefer to forming a bridge structure shown in Figure 4 (a) and 4(c). The O* species prefer to adsorb on the V site no mater we put it on Bi or V stie initially. This is not surprising because the V sites and Bi sites are close to each other, and the distance of V-Bi is only ~3.8 Å. There are 1 d electron on the V atoms which makes it more adsorption than the Bi sties. The free energy diagram of OER is shown in Figure 5. The first proton removal step is very easy due to the bridge adsorption. The second step is also very easy. But the third step is quite difficult which indicates the O-O bond is difficult to form. The rate-determining step is the third step and the overpotential is 2.22 V. Water oxidation cannot occur on the pure BVOF surface due to the large overpotential.

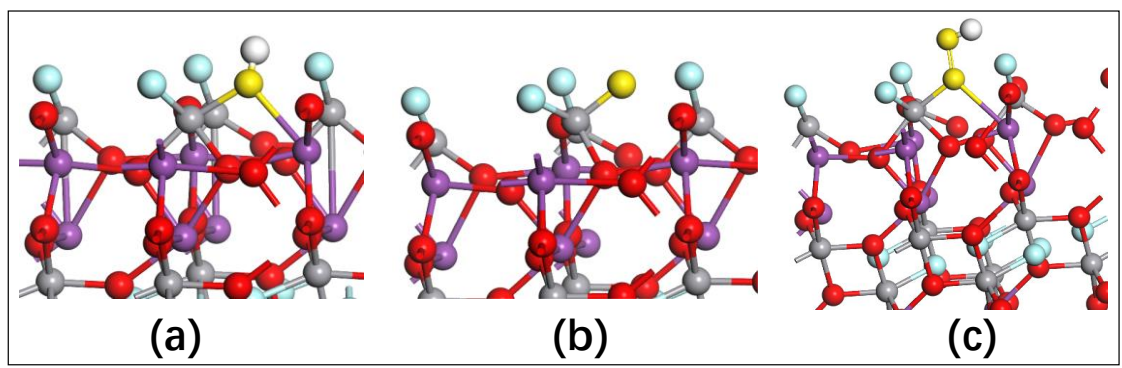

Figure 4. The structures of OER intermediates on (001) surface of BVOF. The yellow spheres are the oxygen atoms form water, and the white spheres are the hydrogen atoms.

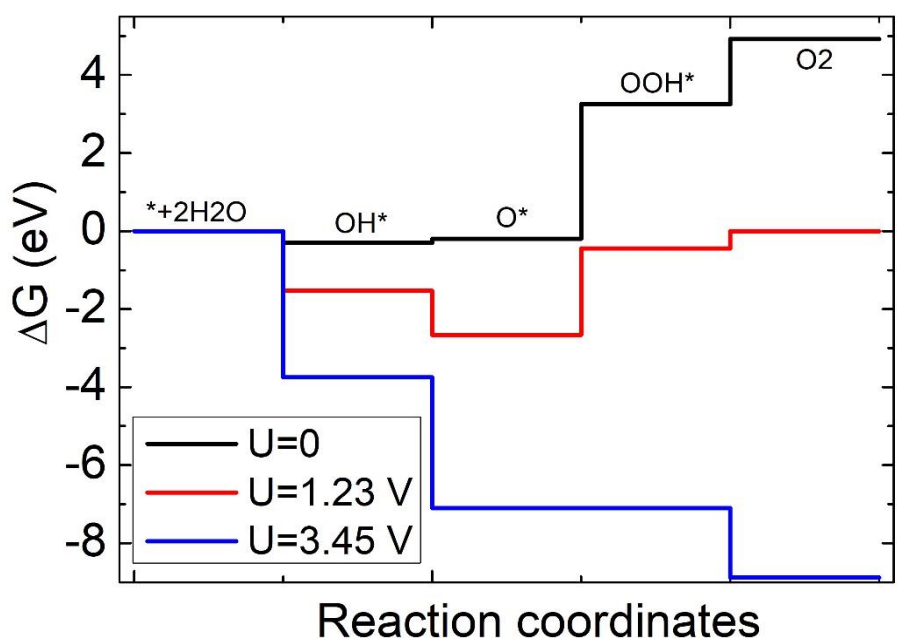

Figure 5. Free energy diagram of OER on (001) surface of BVOF.

4. Conclusion

In this work, the structural, electronic, and surface reaction properties of BVOF were investigated using PBE and hybrid functionals. The calculated bandgap is close to experimental result only using the hybrid functional with a fraction of HF exchange α around 0.1. The direct V 3d to V 3d excitation make the bandgap lower which are benefit for the light absorption but may be not an advantage for the charge separation as electrons and holes are all generated on the V atoms. The (001) surface is the most stable among all the low index (001), (010), and (100) surfaces. There are both V and Bi sites on the termination of (001) surface. The OER intermediates OH* and OOH* prefer to adsorb on the both V and Bi sites which makes the O-O bond difficult to form. This leads the overpotential of OER is very high, and it is bad for water oxidation on the pure BVOF. Our results show the pure BVOF maybe not a good photocatalyst, and the modification are needed to improve its efficiency in the future.

5. ASSOCIATED CONTENT

Supporting Information

The supporting information includes: phonon frequencies of BVOF at gamma point; the electronic structure of BVOF with K-points $2 \times 2 \times 1$ and $4 \times 4 \times 2$; the band structure of BVOF.

**AUTHOR INFORMATION**

*Corresponding Author:*

Taifeng Liu: tfliu@vip.henu.edu.cn

**Notes**
The authors declare no competing financial interest.

ACKNOWLEDGMENT

This work was supported in part by the National Natural Science Foundation of China (grant # 21703054).

TOC

Oxygen evolution reaction (OER) on (100) surface of BVOF with the overpotential 2.22 V.

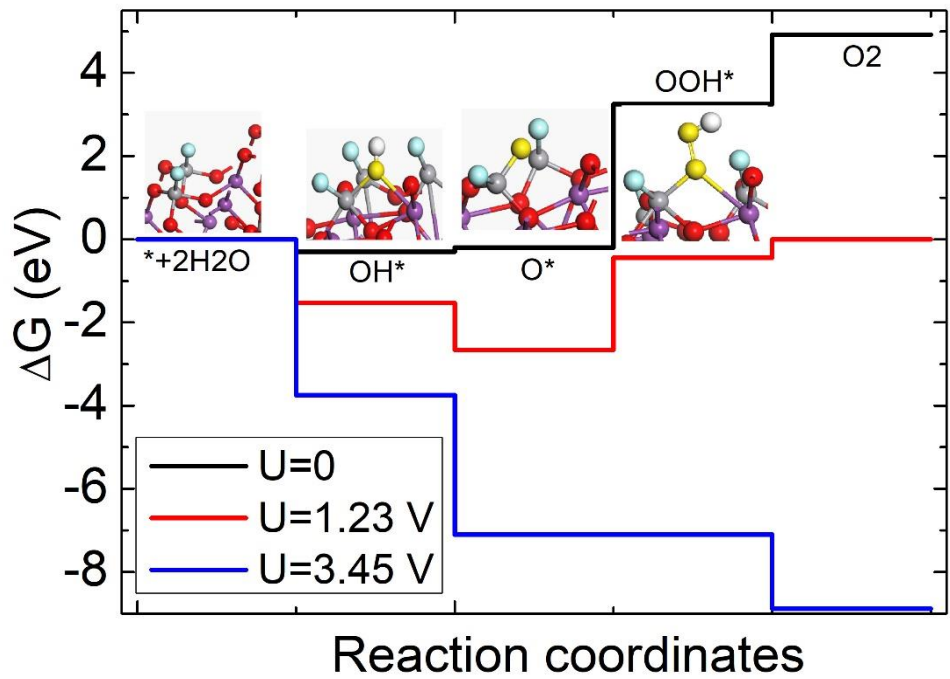